\documentclass{ws-p8-50x6-00}

\begin{document}
\def\today{July 21, 2000}
\newcommand{\beq}{\begin{equation}}
\newcommand{\eeq}{\end{equation}}
\def\be{\begin{eqnarray}}
\def\ee{\end{eqnarray}}
\newcommand{\ben}{\begin{eqnarray*}}
\newcommand{\een}{\end{eqnarray*}}

\def\simlt{\stackrel{<}{{}_\sim}}
\def\simgt{\stackrel{>}{{}_\sim}}
\newcommand{\lnablasq}{\overleftarrow{\nabla}{}^{\!2}}
\newcommand{\rnablasq}{\overrightarrow{\nabla}{}^{\!2}}
\newcommand{\wt}{\widetilde}
\def\({\left( }
\def\){\right) }
\def\[{\left[ }
\def\]{\right] }

\newcommand{\Real}{\mbox{Re}\,}
\newcommand{\Imag}{\mbox{Im}\,}
\newcommand{\order}{{\cal O}}

\newcommand{\drms}{DR[$\overline{\mbox{MS}}$]}
\newcommand{\Mbar}{\frac{M}{4\pi}(\mu+ip)}
\newcommand{\sing}{$^1\!S_0$ }
\newcommand{\trip}{$^3\!S_1$ }
\newcommand{\psing}{$^1\!P_1$ }
\newcommand{\LambdaNN}{\Lambda_{{\rm NN}}}

\newcommand{\dsl}{\raise.15ex\hbox{$/$}\kern-.57em\hbox{$\partial$}}
\newcommand{\Dsl}{\raise.15ex\hbox{$/$}\kern-.65em\hbox{$D$}}
\newcommand{\Tr}{\mbox{Tr}}

\title{Can Merons Describe Confinement?}

\author{James V.\ Steele}

\address{Center for Theoretical Physics and Laboratory for Nuclear
Science\\Massachusetts Institute of
Technology, Cambridge, MA\ \ 02139, USA\\E-mail: jsteele@mit.edu}

\maketitle

\abstracts{
Merons, conjectured as a semiclassical mechanism for color
confinement in QCD, are topological charge-$1/2$, singular solutions
to the classical Yang-Mills equations of motion.
I will discuss how lattice techniques can extend the study of
merons to nonsingular stationary solutions without destroying
properties believed to be essential for confinement, 
and how zero modes can be used
to identify these gauge field configurations in stochastic evaluations
of the lattice QCD path integral.
}

\section{Introduction}

Merons are one of the earliest proposed semiclassical mechanisms
for confinement,~\cite{Callan:1977qs}
but have received less attention than other
approaches~\cite{Gonzalez-Arroyo:1996zy,Faber:1999wq} 
due to analytical limitations.
I will present a new direction for meron studies,%
\footnote{Talk presented at the ``Fourth workshop on Continuous
Advances in QCD,'' Minneapolis, MN, May 12-14, 2000.
The work presented in this talk was done in collaboration with
J.~W.~Negele.~\protect\cite{Steele:2000xk}}
developed recently,~\cite{Steele:2000xk} 
in which lattice gauge theory will be able to 
determine the presence and role
of merons in QCD.

Chiral symmetry breaking and confinement of color charge,
the two essential features of low-energy QCD, cannot be understood
perturbatively in the coupling constant.
Two complementary nonperturbative approaches are 
analytical approximation using
semiclassical analysis of the QCD path integral
(analogous to the WKB approximation in quantum mechanics) 
and numerical solution using lattice gauge theory.
Combining physical understanding from 
the semiclassical approximation
with systematic rigor of the lattice 
has provided valuable insight into hadron structure~\cite{Chu:1994vi}
and can be extended to study confinement.

Most semiclassical treatments of QCD have focused on instantons,
leading to a qualitative, and in some cases even quantitative,
understanding of chiral symmetry breaking and the structure of the QCD
vacuum.~\cite{Schafer:1998wv}
However, simple configurations of instantons, such as a dilute
gas~\cite{Callan:1977qs} or a random superposition,~\cite{Chen:1999ct}
do not confine color charges.
It is possible the complicated dynamics of confinement
escapes any explanation save a full solution of QCD at large
coupling.
However, analogy with lower dimensional models seems to indicate
otherwise.

In 2+1 dimensions, the Georgi-Glashow model, which couples a triplet
Higgs to an SU(2) vector boson field,
contains magnetic monopoles as solutions to the
Euclidean classical equations of motion.
Taking the Higgs to have an expectation value along the third isospin
direction $\phi_a = \delta_{a3}\, v$, the residual massless photon
exhibits a monopole solution of form $A^3_\mu \sim \epsilon_{3\mu\nu}\,
x_\nu/x^2$ at large distances.
Polyakov showed these monopoles condense, leading to confinement of
electric charge.~\cite{Polyakov:1977fu}
Similarly, the Schwinger model (QED in 1+1 dimensions) exhibits
confinement through the dominance of vortex
configurations~\cite{Nielsen:1977jv} of form $A_\mu = \epsilon_{\mu\nu}
x_\nu/x^2$.
These vortices have a topological charge of $\frac12$, 
which also shows up in the quark correlator at finite 
temperature.~\cite{Steele:1995gf}

These semiclassical examples of confinement are intriguing
in that the effect is generated by the same gauge-field configuration
interpreted in different dimensions. 
The meron is the (3+1)-dimensional nonabelian 
extension of the 't~Hooft--Polyakov monopole
configuration with $A^a_\mu = \eta_{a\mu\nu} x_\nu/x^2$.
The important property shared by each of these configurations is 
$A\sim 1/r$ and is not pure gauge.
Then
the integral $\oint\! A_\mu dx^\mu$ over a loop of area $R\times T$ will have
a contribution of order unity as long as the source of the field is within 
$R$ of this loop in any direction.~\cite{Callan:1977qs}
If the vacuum is made up of a uniform distribution of these
semiclassical configurations, the energy between charged particles 
will behave like $E(R)\propto R^d$ in $d$+1 dimensions.
Interactions between monopoles in 2+1 dimensions reduce this down to a
linearly rising potential~\cite{Polyakov:1977fu} and it is 
conjectured that merons
will do the same in 3+1 dimensions.~\cite{Callan:1977qs}

However, a purely analytic study of merons has proven intractable
for several reasons.
Unlike instantons, no exact solution exists for more than two
merons,~\cite{Actor:1979in}
gauge fields for isolated merons fall off too slowly ($A\sim1/x$)
to superpose them, and the field strength is singular.
A patched {\it Ansatz} configuration that removes the
singularities~\cite{Callan:1977qs} does not satisfy the classical
Yang-Mills equations, preventing even
calculation of 
gaussian fluctuations around 
a meron pair.~\cite{Laughton:1980vv}

The properties of merons are reviewed in Sec.~{\bf 2}.
The discussion is restricted to SU(2) color, since topological
objects live in embedded SU(2) subgroups of larger gauge groups.
As work with instantons has shown, fermionic zero
modes are a powerful tool for identifying topological objects on the
lattice.
Therefore, in Sec.~{\bf 3}, the zero mode of the continuum
meron pair {\it Ansatz} is analytically derived.
In Sec.~{\bf 4}, 
a smooth, finite action configuration
corresponding to a meron pair is presented, which with its associated
quantum fluctuations, should be present in 
Monte Carlo lattice QCD calculations.
These exact numerical results along with the associated zero mode
are compared with analytical results of the meron pair {\it Ansatz}.

\section{Analytic Meron Review}

Two known solutions to the classical Yang-Mills
equations in four Euclidean dimensions
are instantons and merons.
Both have topological charge, can be interpreted as tunneling
solutions, and can be written in the general form 
(for the covariant derivative $D_\mu = \partial_\mu - i A_\mu^a
\sigma^a/2$)
\beq
A_\mu^a(x) =  \frac{2\,\eta_{a\mu\nu} \, x^\nu}{x^2} f(x^2)
\ ,
\label{general}
\eeq
with $f(x^2) = x^2/(x^2+\rho^2)$ for an instanton
and $f(x^2)=\frac12$ for a meron.

Conformal symmetry of the classical Yang-Mills action, in
particular under inversion $x_\mu\to x_\mu/x^2$, shows that
in addition to a meron at the origin, there is 
a second meron at infinity. 
These two merons can be mapped to arbitrary positions, defined 
to be the origin and $d_\mu$, by the conformal transformation%
\footnote{A more general conformal transformation only differs
from this choice by either a rotation, a translation, or a dilatation.}
\beq
x_\mu \;\; \to \;\;
d_\mu + d^2 \frac{(x-d)_\mu}{(x-d)^2} \ .
\eeq
After a gauge transformation, the gauge field for the two merons
takes the simple form~\cite{deAlfaro:1976qz}
\beq
A_\mu^a(x) =  \eta_{a\mu\nu} 
\[ \frac{x^\nu}{x^2}+ \frac{(x-d)^\nu}{(x-d)^2} \] \ .
\label{conform}
\eeq
Similar to instantons,~\cite{Schafer:1998wv} a meron pair can be
expressed in 
singular gauge by performing a large gauge transformation
about the mid-point of the pair,~\cite{Hands:1990bu}
resulting in a gauge field
that falls off faster at large distances ($A\sim x^{-3}$).
A careful treatment of the singularities shows that the
topological charge density is~\cite{deAlfaro:1976qz}
\beq
Q(x) \equiv 
\frac{\Tr}{16\pi^2} 
 \( F_{\mu\nu} \wt F^{\mu\nu} \)
 = \frac12\, \delta^4(x) + \frac12\, \delta^4(x-d) \ ,
\label{topch}
\eeq
yielding total 
topological charge $Q=1$, just like the instanton.

The gauge field Eq.~(\ref{conform}) has infinite action density 
at the singularities $x_\mu=\{0,\, d_\mu\}$.
Hence, a finite action {\it Ansatz} has been suggested~\cite{Callan:1977qs}
\beq
A_\mu^a(x) =  \eta_{a\mu\nu} x^\nu \left\{ 
\begin{array}{cll}
\displaystyle
\frac{2}{x^2+r^2} \ ,\qquad & \sqrt{x^2} < r \ ,
\qquad & \mbox{region I,}
\\[3ex]
\displaystyle
\frac{1}{x^2} \ ,\qquad & r < \sqrt{x^2} < R \ ,
\qquad & \mbox{region II,}
\\[3ex]
\displaystyle
\frac{2}{x^2+R^2} \ ,\qquad & R < \sqrt{x^2} \ ,
\qquad & \mbox{region III,}
\end{array}
\right.
\label{caps}
\eeq 
with arbitrary radii $r$ and $R$.
Here, the singular meron fields for $\sqrt{x^2}<r$ and $\sqrt{x^2}>R$
are replaced by instanton caps, each containing topological charge 
$\frac12$ to agree with Eq.~(\ref{topch}).
The action for this configuration is
\beq
S= \frac{8\pi^2}{g^2} + \frac{3\pi^2}{g^2} \ln \frac{R}{r} \ ,
\label{action}
\eeq
which shows the divergence in the $r\to0$ or $R\to\infty$ limit.
There is no angular dependence in this patching, and so
the conformal symmetry of the meron pair is retained.
For example, under a dilatation $x_\mu\to \lambda\, x_\mu$,
both $r$ and $R$ get multiplied by $1/\lambda$ but the ratio and hence
the action Eq.~(\ref{action}) remains invariant.
Although this patching of instanton caps is continuous,
the derivatives are not, and so the equations of
motion are violated at the
boundaries of the regions in Eq.~(\ref{caps}).

\begin{figure}[b]
\epsfxsize=3.75in
\centerline{\epsfbox[170 608 460 720]{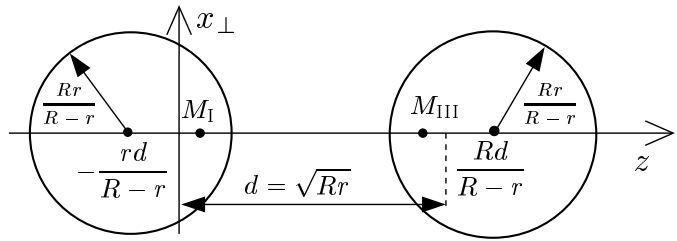}}
\caption{\label{fig1} Meron pair separated by $d=\sqrt{R r}$ regulated
with instanton caps, each containing $\frac12$ topological charge.
} 
\end{figure}

Applying the same transformations used to attain Eq.~(\ref{conform})
to the instanton cap solution, regions I and III become
four-dimensional spheres 
each containing half an instanton.~\cite{Callan:1977qs}
The geometry of the instanton caps
is shown in Fig.~\ref{fig1} for the symmetric choice
of displacement $d=\sqrt{R r}$ along the $z$-direction.
Note that the action for the complicated geometry of
Fig.~\ref{fig1} is still given by Eq.~(\ref{action}), 
which for $d=\sqrt{R r}$ can be rewritten as
\beq
S = S_0 \( 1+ \frac{3}{4} \ln \frac{d}{r} \) \ ,
\label{action2}
\eeq
with $S_0 = 8\pi^2/g^2$.
This case will be used below, and generalization to a different
$d$ is straightforward.

The original positions of the two merons $x_\mu=\{0,\,d_\mu\}$
are not the centers of the spheres, nor are they the positions of
maximum action density, which occurs within the spheres  
with $S_{\rm max} = (48/g^2) (R+r)^4/d^8$ at
\beq
\(M_{\rm I}\)_\mu = \frac{r^2}{r^2+d^2} \; d_\mu \ ,
\qquad
\(M_{\rm III}\)_\mu = \frac{R^2}{R^2+d^2} \; d_\mu \ . 
\label{maxima}
\eeq
However, in the limit $r\to0$ and
$R\to\infty$ holding $d$ fixed, the spheres will shrink around the
original points reducing to the bare meron pair in Eq.~(\ref{conform}). 
In the opposite limit $R\to r$, the radii of the spheres
increase to infinity, leaving an instanton of size $\rho = d$.

An instanton can therefore be interpreted as consisting of a meron
pair. 
This complements the fact that instantons and antiinstantons have
dipole interactions between each other.
If there exists a regime in QCD where meron entropy contributes more
to the free energy than the logarithmic potential between pairs, like
the Kosterlitz--Thouless phase transition, 
instantons will break apart into meron
pairs,~\cite{Callan:1977qs}
with the intrinsic size of the instanton caps determined by the
original instanton scale $\rho$.  
This will occur when the entropy to create meron pairs, 
which is $R/r$ in each space-time direction
with the general scales $r$ and $R$ given in Eq.~(\ref{caps}),
overcomes the logarithmic interaction energy Eq.~(\ref{action})
to dominate the QCD path integral, 
or schematically,~\cite{Callan:1977qs}
\beq
\(\frac{R}{r}\)^4 e^{-3\pi^2 \ln (R/r) /g^2(s)}  = 
\(\frac{R}{r} \)^{4- 3\pi^2/g^2(s)}
\quad\simgt\quad 1
\ .
\label{entropy}
\eeq
How the scale $s$ in the renormalized coupling
constant depends on $r$ and $R$ is not known.
It is assumed that as longer and longer distances are probed, $s$
also increases, causing entropy to win out over energy in
Eq.~(\ref{entropy}).
This occurs for relatively weak coupling, $g^2(s)/8\pi^2 >
\frac{3}{32}$, so the semiclassical
approximation should still be valid.~\cite{Callan:1977qs}
Lattice calculations can test this physical picture
by increasing the coupling constant and seeing if 
the predominant semiclassical degrees of freedom change from
a dilute gas of instantons into a plasma
of merons.

\section{Analytic Zero Modes}

The Atiyah-Singer index theorem states that 
a fermion in the presence of 
a gauge field with topological charge $Q$, described by the equation
\beq
\Dsl\, \psi = \lambda \psi \ , \quad \mbox{with} \quad
\psi = \( {\psi_R \atop \psi_L} \) \ ,
\label{dirac}
\eeq
has $n_R$ right-handed and $n_L$ left-handed
zero modes (defined by $\lambda=0$) such that $Q=n_L-n_R$.
This can sometimes be
strengthened by a vanishing theorem.~\cite{Jackiw:1977pu}
Applying $\Dsl$ twice to $\psi$ decouples the right- and left-hand
components, and focusing on the equation for $\psi_R$, gives
\beq
\( D^2 + \frac12 \bar\eta_{a\mu\nu} \sigma^a F^{\mu\nu} \) \psi_R 
=0\ ,
\label{van}
\eeq
where $\bar\eta_{a\mu\nu}$ denotes the 't~Hooft symbol, $\sigma^a$
acts on the spin indices of $\psi_R$, and $F^{\mu\nu}$ 
acts on the color indices.
For a self-dual gauge field (like an instanton), the second term in
Eq.~(\ref{van}) is zero; and since $D^2$ is a negative definite
operator, there are no 
normalizable right-handed zero modes, implying $Q=n_L$.
Although the meron pair is not self-dual, the second term can be shown
to be negative definite as well, leading to the same conclusion.

In general, a gauge field in Lorentz gauge with $Q=1$ can be
written in the form 
\beq
A_\mu^a(x) = - \eta_{a\mu\nu} \partial_\nu \ln \Pi(x) \ . 
\label{pi}
\eeq
This has a fermion zero mode given by
\beq
\psi = \( {0 \atop \phi} \) \ , \qquad
\phi_\alpha^a = {\cal N} \;  \Pi^{3/2} \;
\varepsilon_\alpha^a \ ,
\label{zmode}
\eeq
with normalization ${\cal N}$ and
$\varepsilon = i \sigma_2$ coupling the spin index $\alpha$ to the
color index $a$ (both of which can be either 1 or 2) in a singlet
configuration.
The gauge field for the meron pair with instanton caps can be
expressed in the form Eq.~(\ref{pi}) with
\beq
\!\!\!\Pi(x) = \left\{
\begin{array}{ll}
\displaystyle
\frac{2 \xi_i d^2}{x^2 + \xi_i^2 (x-d)^2}\ , 
& \mbox{for regions $i=$ I, III,}
\\[3ex]
\displaystyle
\frac{d^2}{\sqrt{x^2 (x-d)^2}} \ , 
& \mbox{region II,} 
\end{array}
\right.
\eeq
with $\xi_{\rm I}=r/d$ and $\xi_{\rm III}=R/d$.
The normalized solution to the zero mode is then Eq.~(\ref{zmode})
with 
\beq
{\cal N}^{-1} = 2\pi d^2 \[2-\sqrt{\frac{r}{R}}\]^{1/2}
\ .
\eeq
Note that the unregulated meron pair ($r\to0$, $R\to\infty$)
has a normalizable zero mode itself~\cite{Konishi:1999re} 
\beq
\phi_\alpha^a(x) = \frac{d\; \varepsilon_\alpha^a}
{2\sqrt{2} \pi \( x^2 (x-d)^2 \)^{3/4}} \ .
\eeq
The gauge-invariant zero mode density $\psi^\dagger \psi(x)$ 
has a bridge between two merons that falls off like $x^{-3}$
in contrast to the $x^{-6}$ fall-off in all other directions.
This behavior can be used to identify merons when analyzing their zero
modes on the lattice, similar to what was done for instantons in
Ref.~[16].

\section{Merons on the Lattice}

As mentioned above, the patching of instanton caps to obtain an
explicit analytic solution has unavoidable and unphysical
discontinuities in the action density.
Therefore, 
in order to study this solution further,
the gauge field is put on a space-time lattice of spacing $a$ in a
box of size $L_0\times L_1\times L_2\times L_3$.
The gauge-field degrees of freedom
are replaced by the usual parallel transport link variable,
\beq
U_\mu(x) = {\rm P} \exp \[-i \int_x^{x+a\, \hat e_\mu} 
A_\nu(z) dz^\nu\] \ .
\eeq
The exponentiated integral can be performed analytically
within the instanton caps, producing arctangents.
Outside of the caps, 
the integral is evaluated numerically by 
dividing each link
into as many sub-links as necessary to reduce the 
$\order(a^3)$ path-ordering errors below machine precision.

\begin{figure}
\epsfxsize=4.4in
\centerline{\epsfbox{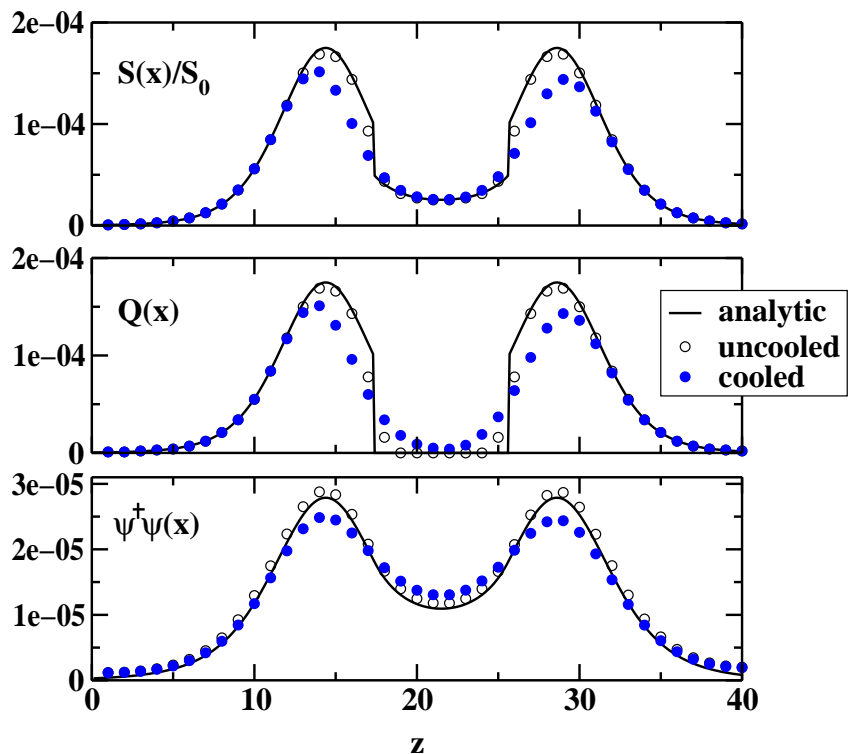}}
\caption{\label{fig2} The action density $S(x)$ for the
regulated meron pair,
normalized by the instanton action $S_0=8\pi^2/g^2$, 
sliced through the center of the configuration in the direction of
separation. 
Shown are the analytic (solid), initial lattice (open circles)
and cooled lattice (filled circles) configurations.  The same is shown
for the topological charge density $Q(x)$ and fermion zero mode
density $\psi^\dagger\psi(x)$.}
\end{figure}

\begin{figure}[t]
\epsfxsize=3.4in
\centerline{\epsfbox{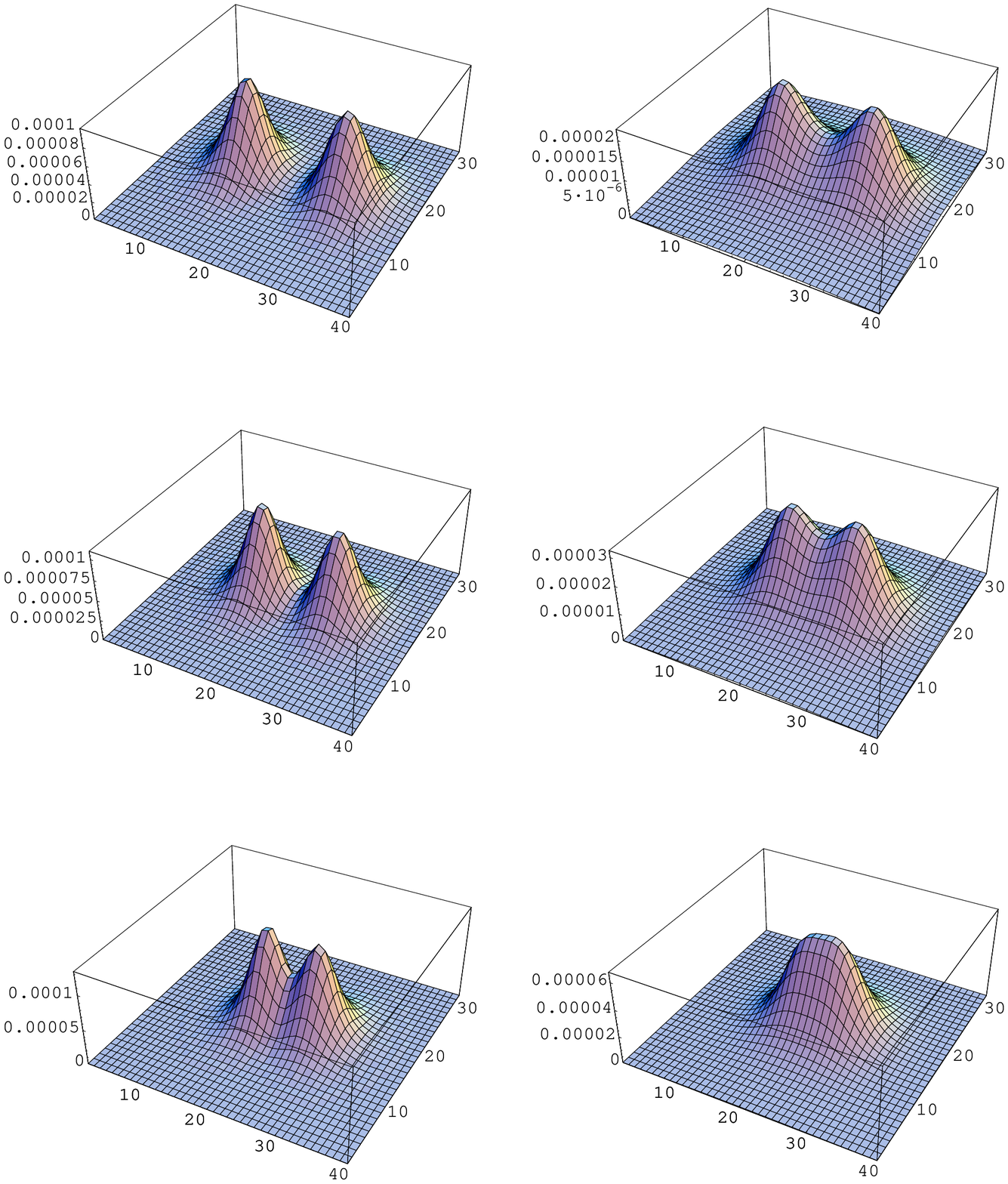}}
\caption{\label{fig3} The topological charge density $Q(x)$ on the
left and
fermion zero mode density $\psi^\dagger\psi(x)$ on the right
for a cooled meron pair
with $r=9$, separated by distance $d=20.8$, $17.7$, and $14.1$ respectively,
in the $(z,t)$-plane. }
\end{figure}

Calculating the action density $S(x)$ for a meron pair with
instanton caps using the improved action of
Ref.~[17] 
and the topological charge density $Q(x)$ using 
products of clovers, the lattice results (open circles) are
compared with the patched {\it Ansatz}
results (solid line)
in Fig.~\ref{fig2} for $r=9$ and $d=20.8$ 
(in units of the lattice spacing) on a $32^3\times40$ lattice.
The Arnoldi method is used to solve for the zero mode of this
gauge configuration and hence the density $\psi^\dagger \psi(x)$,
which is also compared with the analytic result in the same figure,
showing excellent agreement.

Two important features of the patched {\it Ansatz} also evident with the
lattice representation are the discontinuities in the action density 
at the boundary of the instanton caps and
vanishing of the topological charge density outside of the caps as given
by Eq.~(\ref{topch}).
The discontinuity in the action density is unphysical and can be 
smoothed out by using a
relaxation algorithm to iteratively minimize the lattice
action.~\cite{Berg:1981nw}
On a sweep through the lattice, referred to as a cooling step,
each link is chosen to locally minimize the action density.
Since a single instanton is already a minimum of the action, 
this algorithm would leave the instanton unchanged (for a suitably
improved lattice action).
The result for the regulated meron pair
after ten cooling steps is represented in Fig.~\ref{fig2} by the
filled circles, showing the discontinuities in the action density
have already been smoothed out.
Figure~\ref{fig3} shows side-by-side the cooled $Q(x)$ and
$\psi^\dagger \psi(x)$  
in the $(z,t)$-plane for three different meron pair separations.
As the separation vanishes, the zero mode
goes over into the well-known instanton zero mode result.

Like instanton-antiinstanton pairs, however, 
a meron pair is not a strict minimum of the action,
since it has a weak attractive interaction Eq.~(\ref{action})
and under repeated relaxation will coalesce to form an
instanton. 
This is analogous to the annihilation of an instanton-antiinstanton
pair through cooling.~\cite{Chu:1994vi}
Nevertheless, just as it is important to sum all the quasi-stationary
instanton-antiinstanton configurations to obtain essential
nonperturbative physics,~\cite{coleman}
meron pairs may be expected to
play an analogous role.
A precise framework for including quasi-stationary meron pairs is to
introduce a constraint ${\cal Q}[A]$ on a suitable collective variable
$q$ (here chosen to be the quadrupole moment $3z^2-x^2-y^2-t^2$ of
the topological charge)  
as follows
\begin{eqnarray}
Z &=& \int\!\! DA \; \exp \left\{-S[A]\right\} = \int\!\! dq \; Z_q \ ,
\\
Z_q &=& \int\!\! DA \; \exp \left\{ -S[A] - \lambda \( {\cal Q}[A]-q
\)^2 \right\} \ .
\end{eqnarray}
The meron pair is then a true minimum of the effective action with
constraint, allowing for a semiclassical treatment of $Z_q$.
Afterwards, $q$ is integrated to obtain the full partition function $Z$.

\begin{figure}[t]
\epsfxsize=4.5in
\centerline{\epsfbox{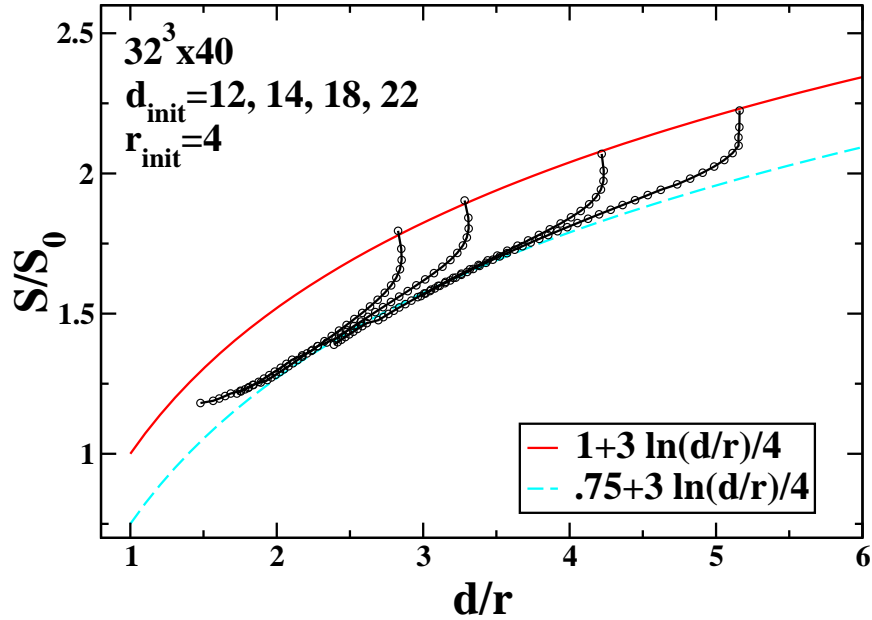}}
\caption{\label{fig4} Action of a regulated meron pair 
as a function of $d/r$ for initial lattice configurations of $d_{\rm
init}=12$, 14, 18, 22 and $r_{\rm init}=4$ on a $32^3\times40$ lattice. 
The circles show cooling trajectories with each cooling step marked.}
\end{figure}

The criterion for a good collective variable $q$ of the system is that
the gradient in the direction of $q$ is small compared to the
curvature associated with all the quantum fluctuations.
In this case, an adiabatic limit results in which 
relaxation of the unconstrained
meron pair slowly evolves through a sequence of
quasi-stationary solutions,
each of which is close to a corresponding stationary constrained solution. 
Detailed comparison of the quasi-stationary and constrained solutions
shows this adiabatic limit is well satisfied.
Therefore, the action of a
meron pair as it freely coalesces into an instanton is presented here.
To compare with the patched {\it Ansatz},  the meron separation $d$
and radii $r$ and $R$ of the lattice configuration are required.
These are found by first measuring the separation of the
two maxima in the action density $\Delta\equiv|M_{\rm I}-M_{\rm III}|$ 
(using cubic splines)
and their values (which are the same in the symmetric case,
denoted by $S_{\rm max}=48 /g^2 w^4$),
and then using Eq.~(\ref{maxima}) to give
\beq
d^2 = \Delta^2 + 4 w^2 \ ,
\qquad
r = \frac{2w d}{d+\Delta} \ ,
\qquad
R = \frac{2w d}{d-\Delta} \ .
\eeq
In Fig.~\ref{fig4}, the total action of a meron pair is plotted as a function
of $d/r$ (solid line), which for the analytic case is given by
Eq.~(\ref{action2}). 
Also shown are cooling trajectories for 
four lattice configurations with different
initial meron pair separations.
Each case starts with a patched {\it Ansatz} of a given separation and the
lattice action agrees with that of the analytic {\it Ansatz}.

The first few cooling steps primarily decrease the action without
changing the collective variable (which is now effectively $d/r$), 
as observed above in Fig.~\ref{fig2}.
Further cooling gradually decreases the collective variable, tracing
out a new
logarithmic curve for the total action
(dashed line in Fig.~\ref{fig4}), 
which is about $0.25$ smaller than the analytic case.
This curve represents the total action of the smooth adiabatic or
constrained lattice meron pair solutions.

The essential property of merons that could allow 
them to dominate the path integral and confine color charge
is the logarithmic interaction Eqs.~(\ref{action},\ref{action2})
which is weak enough to be dominated by the meron 
entropy.~\cite{Callan:1977qs}
Hence, the key physical result from Fig.~\ref{fig4} is the fact that
smooth adiabatic or constrained lattice meron pair solutions clearly
exhibit this logarithmic behavior.

In summary, stationary meron pair solutions have been found on the
lattice exhibiting a logarithmic interaction and 
fermion zero mode localized about the individual merons in agreement
with analytic results. 
The presence and role of merons in numerical evaluations of the
QCD path integral should be investigated, and the study of fermionic
zero modes and cooling of gauge configurations are possible tools
to do so.

\vfill

\section*{Acknowledgments}
I would like to thank the organizers, K.~Olive, A.~Vainshtein,
M.~Shifman, and M.~Voloshin, for an enjoyable and informative workshop.
This work was supported in part by the U.S. Department of Energy under
cooperative research agreement \#DF-FC02-94ER40818.



\begin{thebibliography}{99}

\bibitem{Callan:1977qs}
C.~G.~Callan, R.~Dashen and D.~J.~Gross,
Phys.\ Lett.\  {\bf B66}, 375 (1977);
%
Phys.\ Rev.\  {\bf D17}, 2717 (1978);
%
Phys.\ Rev.\  {\bf D19}, 1826 (1979).
%
\bibitem{Gonzalez-Arroyo:1996zy}
A.~Gonzalez-Arroyo and P.~Martinez,
Nucl.\ Phys.\  {\bf B459}, 337 (1996).
[hep-lat/9507001].
%
\bibitem{Faber:1999wq}
M.~Faber, J.~Greensite, S.~Olejnik and D.~Yamada,
hep-lat/9912002.
%
\bibitem{Steele:2000xk}
J.~V.~Steele and J.~W.~Negele,
hep-lat/0007006.
%
\bibitem{Chu:1994vi}
M.~C.~Chu, J.~M.~Grandy, S.~Huang and J.~W.~Negele,
Phys.\ Rev.\  {\bf D49}, 6039 (1994).
[hep-lat/9312071].
%
\bibitem{Schafer:1998wv}
T.~Sch\"afer and E.~V.~Shuryak,
Rev.\ Mod.\ Phys.\  {\bf 70}, 323 (1998).
[hep-ph/9610451].
%
\bibitem{Chen:1999ct}
D.~Chen, R.~C.~Brower, J.~W.~Negele and E.~Shuryak,
Nucl.\ Phys.\ Proc.\ Suppl.\  {\bf 73}, 512 (1999).
[hep-lat/9809091].
%
\bibitem{Polyakov:1977fu}
A.~M.~Polyakov,
Nucl.\ Phys.\  {\bf B120}, 429 (1977).
%
\bibitem{Nielsen:1977jv}
N.~K.~Nielsen and B.~Schroer,
Phys.\ Lett.\  {\bf B66}, 475 (1977).
%
\bibitem{Steele:1995gf}
J.~V.~Steele, J.~J.~Verbaarschot and I.~Zahed,
Phys.\ Rev.\  {\bf D51}, 5915 (1995)
[hep-th/9407125].
%
\bibitem{Actor:1979in}
A.~Actor,
Rev.\ Mod.\ Phys.\  {\bf 51}, 461 (1979).
%
\bibitem{Laughton:1980vv}
D.~G.~Laughton,
Can.\ J.\ Phys.\  {\bf 58}, 845, 859 (1980).
%
\bibitem{deAlfaro:1976qz}
V.~de Alfaro, S.~Fubini and G.~Furlan,
Phys.\ Lett.\  {\bf B65}, 163 (1976).
%
\bibitem{Hands:1990bu}
S.~Hands,
Nucl.\ Phys.\  {\bf B329}, 205 (1990).
%
\bibitem{Jackiw:1977pu}
R.~Jackiw and C.~Rebbi,
Phys.\ Rev.\  {\bf D16}, 1052 (1977).
%
\bibitem{Konishi:1999re}
K.~Konishi and K.~Takenaga,
hep-th/9911097.
%
\bibitem{Ivanenko:1998nb}
T.~L.~Ivanenko and J.~W.~Negele,
Nucl.\ Phys.\ Proc.\ Suppl.\  {\bf 63}, 504 (1998).
[hep-lat/9709130];
%
T.\ Ivanenko, MIT Ph.D.\ disseration, 1997.
%
\bibitem{GarciaPerez:1994ki}
M.~Garcia Perez, A.~Gonzalez-Arroyo, J.~Snippe and P.~van Baal,
Nucl.\ Phys.\  {\bf B413}, 535 (1994).
[hep-lat/9309009].
%
\bibitem{Berg:1981nw}
B.~Berg,
Phys.\ Lett.\  {\bf B104}, 475 (1981).
%
\bibitem{coleman}
S.~Coleman, {\it Aspects of Symmetry}, (Cambridge University Press, 1988).

\end{thebibliography}
\end{document}